\documentclass[preprint,5p,times,twocolumn]{elsarticle}

\usepackage{amssymb}
\usepackage{amsmath,amssymb}
\usepackage{type1cm}
\usepackage{mathrsfs}
\usepackage{url}
\usepackage{graphicx}
\usepackage{color}
\usepackage{comment}
\newcommand{\argmax}{\mathop{\rm max}\limits}
\newcommand{\argmin}{\mathop{\rm min}\limits}
\newcommand{\figscale}{1.0}
\newcommand\HL[1]{{\color{black}#1}}

\journal{Phys. Rev.E}

\begin{document}

\begin{frontmatter}



\title{
Effect of walking-distance on a queuing system of totally asymmetric simple exclusion process equipped with functions of site assignments
}


\author[RCAST]{Satori Tsuzuki}
\ead{tsuzuki@jamology.rcast.u-tokyo.ac.jp}

\author[RCAST]{Daichi Yanagisawa}
\ead{tDaichi@mail.ecc.u-tokyo.ac.jp}

\author[RCAST]{Katsuhiro Nishinari}
\ead{tknishi@mail.ecc.u-tokyo.ac.jp}

\address[RCAST]{Research Center for Advanced Science and Technology, The University of Tokyo, 4-6-1, Komaba, Meguro-ku, Tokyo 153-8904, Japan}

\begin{abstract}
This paper proposes a totally asymmetric simple exclusion process 
on a traveling lane, which is equipped 
with a queueing system and functions of site assignments along the parking lane.
In the proposed system, new particles arrive at the rear of the queue existing 
at the leftmost site of the system.
A particle at the head of the queue selects one of the empty sites 
in the parking lane and reserves it for stopping at once during its travel.
The arriving time and staying time in the parking sites follow half-normal distributions.
The random selections of empty sites are controlled by the bias of 
the exponential function.
Our simulation results show the unique shape of site usage distributions.
In addition, the number of reserved sites is found to increase 
with an S-shape curve as the bias to the rightmost site increases. 
To describe this phenomena, we propose an approximation model, which 
is derived from the birth-death process and extreme order statistics.
A queueing model that takes the effect of distance from the leftmost site 
of the traveling lane into consideration is further proposed. 
Our approximation model properly describes the distributions of site usage, and the proposed queueing model shows good agreement with the simulation results. 
\end{abstract}


\end{frontmatter}






\section{Introduction}

The queueing theory, which was started by A. K. Erlang\cite{ERLANG-A-K} 
at the beginning of 20th century, has 
attracted many scientists and researchers. 
Most of the theory is still in veil; nevertheless, a strong demand for this theory 
exists not only in academic studies of non-equilibrium statistical physics 
but also in many engineering fields such as
traffic system\cite{RevModPhys.73.1067}, human 
dynamics\cite{PhysRevE.85.021139, PhysRevE.75.026102}, and molecular motor transport\cite{PhysRevE.97.022402, PhysRevE.85.061915}. 
The study of queueing systems has been associated with 
the totally asymmetric simple exclusion process (TASEP) because of two main features:
transportation in a one-way direction and volume exclusion effect, which are
suitable for the simulation of queueing systems\cite{PhysRevE.80.051119, PhysRevE.84.051127, PhysRevE.83.051128}.

This paper proposes a totally asymmetric simple exclusion process 
on a traveling lane, which is equipped with a queueing system and site assignments along the parking lane, under open boundary conditions. In the proposed system, new particles arrive at the rear of the queue existing at the leftmost site of the system. Thereafter, a particle at the head of the queue selects one of the empty sites in the parking lane and reserves it for stopping once during its travel. 
The arriving time and staying time in the parking sites follow half-normal distributions. 
The random selections of empty sites in the site assignments are controlled
by the bias of the exponential function. 

Similar mechanics of the proposed system can be observed in many real-world cases
(e.g. parking problems in highways, airplane boarding, and airport ground transportations).
Therefore, studying the proposed system is meaningful in many application fields. 
In particular, it is important to investigate 
the relationship between the occupancy of parking sites and the ways of site assignments
because they are closely related with each other. 
The major scope of this research is to describe the relationship between 
the effect of site assignments in the proposed system and the occupancy of parking sites.

\HL{Because the parking site can be regarded as an absorption site in a wider sense, 
	the proposed model is classified into the same category of the studies 
		on multiple-lane systems with Langmuir Kinetics.}
Many previous studies have been reported, 
exemplified by \cite{Ezaki2011PositiveCE, Ichiki2016} for parallel-lane systems under periodic conditions, 
\cite{Verma2015} for triple parallel-lane systems with Langmuir Kinetics, and 
 \cite{Ichiki2016, Verma2015, PhysRevE.70.046101, doi:10.7566/JPSJ.85.044001, Yanagisawa2016,
 0295-5075-107-2-20007, RePEc:wsi:ijmpcx:v:18:y:2007:i:09:p:1483-1496} for two parallel-lane 
 systems with Langmuir Kinetics. However, queueing problems are not taken into accounts in these studies.
Regarding the function of site assignments, 
our previous research\cite{PhysRevE.97.042117} is a 
pioneering study on site-assingment for parallel-lane systems; however, 
the problem of queueing was not discussed in that study.

In the modeling of queues, we consider the effect of walking distance from the entrance.
Several previous studies have worked on this problem of walking-distance for specific cases, exemplified by the 
D-Fork system\HL{\cite{10.1007/978-3-540-79992-4_59}} , D-Parallel system\HL{\cite{10.1007/978-3-540-79992-4_59}}, and combinational
queueing system of D-Fork with D-Parallel system\cite{YANAGISAWA2013238}. 
A distinguishing factor of the proposed system compared to these previous systems is that
the parking sites (service sites) push the particle back to the traveling 
lane, whereas in the previous studies, the particles pass through the service 
sites and exit from the opposite side of the traveling lane. 
Since the particle reentering the traveling lane causes delay in the traveling 
lane, the ways of site usage distributions
affect the queueing system. 
Hence, the mechanics in the proposed system are different from those 
of the previous studies.
In this paper, we propose an approximation model to describe the site usage distribution of the proposed system 
on the basis of birth-death process for the spatial direction and
extreme order statistics for the time direction.

The remainder of this paper is structured as follows. 
Section~2 provides a summary of the target system and 
that of the classical $M/M/c$ queueing model. 
Section~3 investigates the dependence of the utilization 
of parking sites on the distribution parameter of the exponential function 
through simulations. 
In Section~4, we propose an approximation model that describes 
the site usage distribution of the proposed system on the basis of the birth-death process and 
extreme statistics. Section~5 summarizes our results and concludes this paper.

\section{Models}
\subsection{Target System}
\begin{figure}[t]
\begin{center}
\vspace{-5.0cm}
\includegraphics[width=\figscale\textwidth, clip, bb= 0 0 2026 1127]{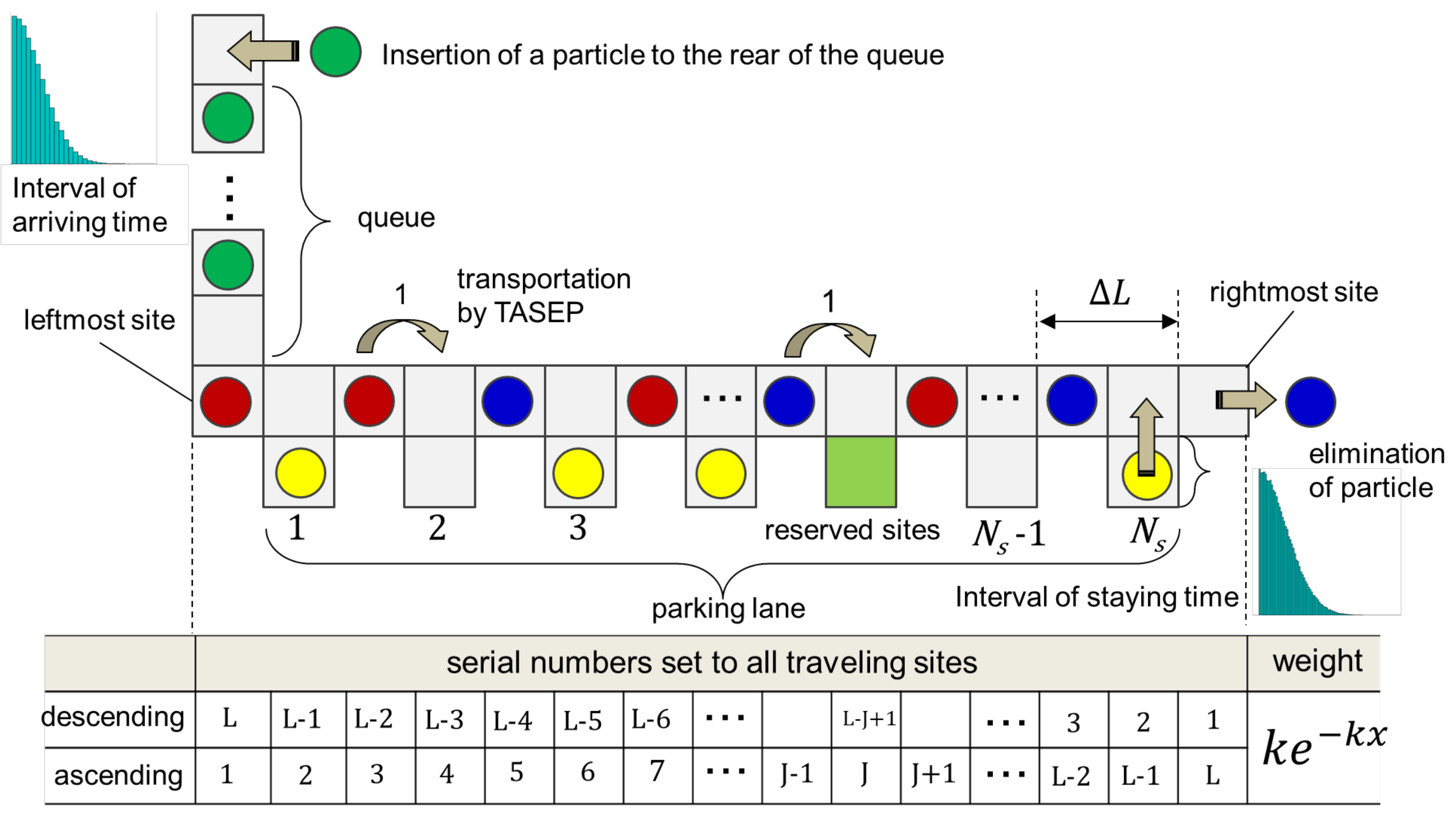}
\end{center}
\caption{Schematic view of the target system.}
\label{fig:schemview}
\end{figure}
A schematic view of the target system is depicted in Fig.\ref{fig:schemview}. 
The system consists of two parallel lanes: 
a traveling lane, which is composed of $L$ sites, and a parking lane, which is composed of $N_{s}$ sites ($N_{s}=L/2$). 
\HL{The distance between two parking sites $\Delta l$ is set to be $2$.}
\HL{In the proposed system, a particle takes four different states. 
The green state indicates that the particle is in a state of queuing. 
The red state indicates that the particle is in transport before stopping at the designated site of the parking lane. 
The yellow state indicates that the particle is currently stopping at the site of the parking lane. 
The blue state indicates that the particle is in transport after exiting the parking lane.
To sum up, the flow of a particle is described as follows. 
A particle arrives at the rear of the queue, which emerges at the leftmost site.
After staying in the parking site, the particle goes back to the traveling lane, changing 
its state from the yellow-colored state to the blue-colored state, and then moves towards the rightmost site.
The particle in the traveling lane is eliminated from the system at the next step after moving to the rightmost site.

Additionally, a parking site takes three kinds of states (reserved state, occupied state, and empty state). 
We call both the occupied state and reserved state simply as ``a busy state'' in this study. 
For the time integration, we adopt the parallel update method.}

\HL{
The interesting subjects of this study are not only the problems with the random arrival 
	such as the parking in highways but also the problems with the scheduled arrival with 
	the random delay such as the airport ground transportations.
	Especially in the latter case, the use of normal distribution is 
	reasonable, however, it has a practical problem 
	in that we have to cut the tail of the left side of the distribution in some cases. 
	In order to avoid this problem, we take up to use the half-normal distribution in this study.
}

\begin{figure}[t]
\begin{center}
\vspace{-7.0cm}
\includegraphics[width=\figscale\textwidth, clip, bb= 0 0 1530 1120]{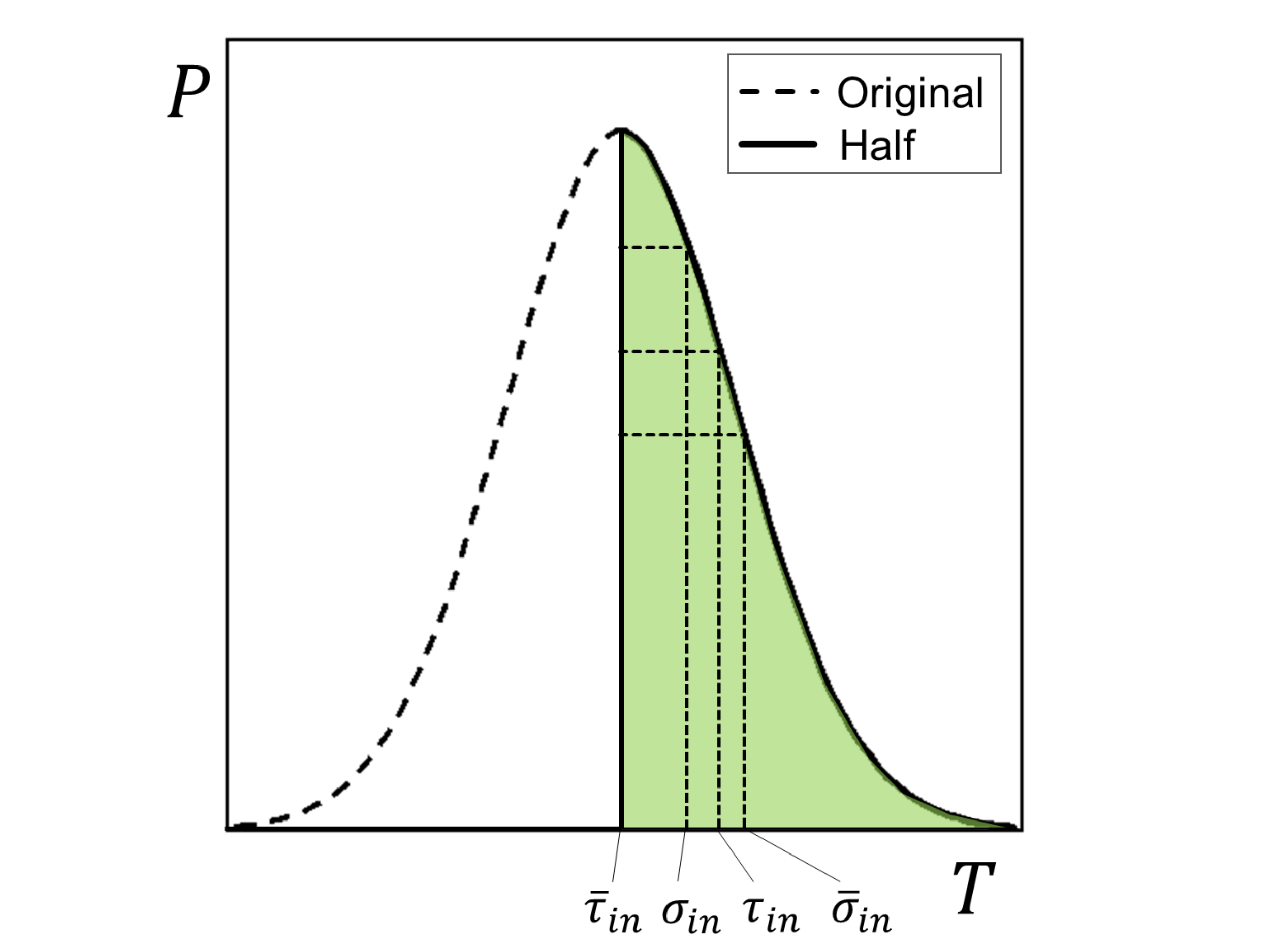}
\end{center}
\caption{\HL{Schematic view of the parameters of the interval of arrival time which follows half-normal distribution.
The horizontal axis indicates the time step and the vertical axis indicates the probability.}}
\label{fig:param:half-normal}
\end{figure}

The interval of the arrival time is set to follow a half-normal distribution;
the mean $\tau_{in}$ and deviation $\sigma_{in}$ of the interval of arrival time
are given as follows:
\begin{eqnarray}
\tau_{in}   &=&  \bar{\tau}_{in} + \bar{\sigma}_{in} \sqrt{\frac{2}{\pi}} \label{eq:half-ndist1}\\ 
\sigma_{in} &=& \HL{\bar{\sigma}_{in} \sqrt{(1-\frac{2}{\pi})}} \label{eq:half-ndist2}
\end{eqnarray}
Here, $\bar{\tau}_{in}$ and $\bar{\sigma}_{in}$ are the mean and deviation 
of the original normal distribution, respectively.
\HL{A schematic view of $\tau_{in}$, $\sigma_{in}$, $\bar{\tau}_{in}$ and $\bar{\sigma}_{in}$ is depicted in Fig.\ref{fig:param:half-normal}.}

A particle at the head of the queue selects one of the empty sites 
in the parking lane and reserves it for stopping once during its travel. 
Similar to that of the arrival time, the interval of the staying time 
is also set to follow a half-normal distribution:
\begin{eqnarray}
\tau_{s}   &=&  \bar{\tau}_{s} + \bar{\sigma}_{s} \sqrt{\frac{2}{\pi}}\label{eq:half-ndist3}\\ 
\sigma_{s} &=& \HL{\bar{\sigma}_{s} \sqrt{(1-\frac{2}{\pi})}} \label{eq:half-ndist4} 
\end{eqnarray}
Here, $\tau_{s}$ and $\sigma_{s}$ are the mean and deviation of the half-normal distribution, while $\bar{\tau}_{s}$ and $\bar{\sigma}_{s}$ are those of 
the original normal distribution, respectively.
Unless otherwise noted, the mean and deviation of the two cases 
are indicated by $\tau_{in}$, $\sigma_{in}$, $\tau_{s}$, and $\sigma_{s}$ in this paper. 


Two important rules are made to the system. 
The particle at the head of the queue is not permitted to reserve the parking 
site that has already been reserved by the other particle unless 
the site releases the particle. 
In addition, the yellow particles in the parking sites have priority access to the upper site on the traveling lane
compared to the red/blue colored particle, which is to access the same \HL{traveling site}.

The random selection of an empty parking site by the particle at the head of the queue 
is controlled by the exponential function $k e^{-k x}~(x \ge 0, k > 0)$. 
In the case of ``ascending'' in Fig.\ref{fig:schemview}, the random selection 
of the empty sites becomes biased toward the leftmost site 
as the parameter increases. 
In the case of ``descending'' in Fig.\ref{fig:schemview}, 
the random selection becomes biased toward the opposite rightmost site 
by setting the reversed sequential number to the number of sites $L$. 
Note that the inverse transform sampling (ITS) \cite{doi:10.1137/1.9780898717570, Lecuyer2011} is introduced to 
generate random variables that follow the exponential distribution.

We denote the latter type of bias by multiplying the negative sign to parameter $k$ for the sake of easy view. 
Namely, in the notation of $k e^{-k x}$, the random selection gets biased towards the leftmost site as parameter $k$ 
increases while $k > 0$. 
In contrast, it gets biased towards the rightmost site as parameter $k$ decreases while $k < 0$.
In the case of setting parameter $k$ to be zero, no external bias is given to the random selection.

\subsection{The classical $M/M/c$ queue}
In this section, we overview the classical queueing theory.
A queueing system is characterized by six stochastic properties: 
the arrival process $A$, service process $B$, number of servers in the system $C$, maximum 
number of possible customers who will arrive at the system $K$, number of sources of customers $N$, and 
service discipline $D$. All these properties are summarized as $A/B/C/K/N/D$ by Kendall's notation\cite{kendall1953}. 
\HL{The notation of $K$ and $N$ are abbreviated in case of infinity and that of $D$ is abbreviated in case of FIFS (First Come First Served); 
in this case, the system can be represented simply as $A/B/C$.}
The proposed system in this study is categorized into 
$M/M/c$ queueing sytems because the arriving time 
and staying time follows Markov process and the system has finite number of parking sites $N_{s}$. 
Note that the left and right $M$ indicate the Markov process, and 
the notation $c$ indicates the number of servers 
(the $c$ corresponds to $N_{s}$ in the system).
In this section, several important formulas of the classical $M/M/c$ queueing system are enumerated.
For more details on queueing theories, refer to\cite{doi:10.1002/0471200581.ch7, Gross:2008:FQT:1972549}

\HL{The arrival rate $\lambda$ and service rate $\mu$ are defined as the characteristic values of the queueing system.
On the condition that the $\lambda$ and $\mu$ are given as constant parameters,}
a distribution of probability $P_{n}$ that the whole system (including queue) has $n$ customers 
at a stationary state is obtained as a consequence of solving the transition equation of length of queue between the time step $n$ and time step $n+1$. 
\begin{eqnarray}
P_{n} &=& 
\begin{cases}
\frac{a^{n}}{n!} P_{0}~~~~~~~~(n=1,2,\dots, c)\\
\frac{a^{n}}{c^{n-c}c!} P_{0}~~~~(n=c+1, c+2, \dots) \label{eq:mmc:prob}
\end{cases} \\
P_{0} &=& \Biggl\{ \sum_{n=0}^{c-1} \frac{a^{n}}{n!} + \frac{a^{c}}{c!}\frac{1}{1-\rho}\Biggr\}^{-1} \\
a &=&\frac{\lambda}{\mu} \label{eq:mmc:a} \\
\rho&=&\frac{\lambda}{c\mu}\label{eq:mmc:mu}
\end{eqnarray}
Additionally, the length of queue $L_{q}$, total number of customers 
in the whole system $L$, and the number of 
utilized servers $U$ are obtained from Eq.(\ref{eq:mmc:prob}) to Eq.(\ref{eq:mmc:mu}), as follows: 
\begin{eqnarray}
L_{q} &=& \sum_{n=c+1}^{\infty}(n-c) P_{n} = C(c,a)\frac{a}{c-a} \\
C(c,a) &=& \frac{c}{c-a}\frac{a^{c}}{c!} P_{0} \\
L     &=& \sum_{n=0}^{\infty}nP_{n} = L_{q}+a \\
U     &=& L-L_{q} = a \label{eq:mmc:U}
\end{eqnarray}
Consequently, the utilization of servers corresponds to the parameter $a$, which 
is defined as the value of arrival rate divided by service rate, as shown in Eq.(\ref{eq:mmc:a}). 

In the case of $c=1$ ($M/M/1$ queue), the arrival rate $\lambda$ 
and service rate $\mu$ are defined as 
the inversed values of arrival time and service time, as follows:
\begin{eqnarray}
\lambda &=& {\tau_{in}}^{-1} \label{eq:mm1:lam}\\
\mu &=& {\tau_{s}}^{-1} \label{eq:mm1:mu}
\end{eqnarray}
Even in general cases of $c>1$, all the servers are assumed to have same 
the values of arrival rate $\lambda $ 
and service rate $\mu$ defined by Eq.(\ref{eq:mm1:lam}) and 
Eq.(\ref{eq:mm1:mu}), respectively, similar to the $M/M/1$ queue 
in the classical $M/M/c$ queueing model.
However, this assumption causes a non-negligible deviation from 
real-world systems because 
the effect of walking distance to each server is not considered in the classical queueing theory. 
In particular, this assumption becomes a serious problem in the proposed system because the reentering customer causes a delay in transportation on the traveling lane. 
To solve this problem, in this paper, approximation models 
that consider the effect of walking distance are proposed in Section~4.

\begin{figure}[t]
\begin{center}
\vspace{-7.0cm}
\includegraphics[width=\figscale\textwidth,clip, bb= 0 0 1500 1125]{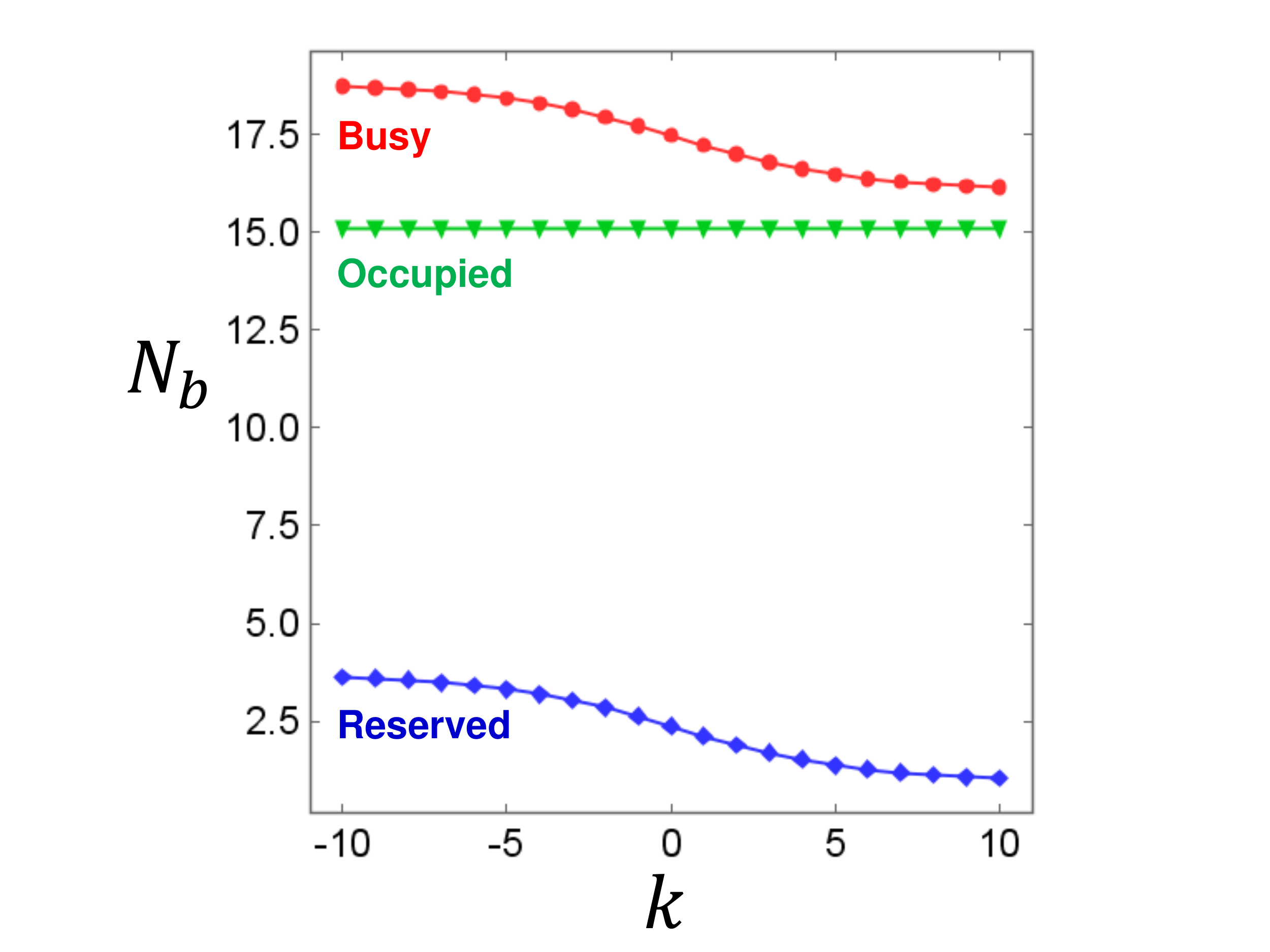}
\end{center}
\caption{Dependence of the number of reserved/occupied sites and the number of busy sites (the sum of reserved sites and occupied sites) 
	on the different values of distribution parameter $k$ between -10 and 10. }
\label{fig:breakdown}
\end{figure}

\section{Simulations}
\HL{
We set the $\tau_{in}$, $\sigma_{in}$, $\tau_{s}$, and 
	$\sigma_{s}$ to be $20$, $1$, $300$ and $10$, respectively, in this paper.}
The particle flux at the rightmost site was measured in proportion to 
the number of parking sites and the number of total time steps 
to determine the condition of reaching the stationary state.
Thereafter, we set the total number of parking sites $N_s$ 
to be $48$ and set the number of total time steps to be $2,000,000$.

Figure~\ref{fig:breakdown} shows the dependence of the number of reserved/occupied sites and the number of busy sites (the sum of reserved sites and occupied sites) 
on the different values of distribution parameter $k$ between -10 and 10.
It was observed that the number of busy sites $N_b$ increases 
with a gentle S-shape curve 
as the parameter $k$ decreases. Substantially, the increase 
in the number of busy sites $N_b$ is found to be determined only
by the increase in reserved sites $N_{r}$.
On the other hand, the number of occupied sites $N_{o}$ 
remained constant during the simulations. 
The result in Fig.\ref{fig:breakdown} indicates that the 
utilization of $U$ servers in the classical M/M/c queueing theory 
corresponds not to the number of occupied sites $N_{o}$, but to the number of busy sites $N_{b}$ in the proposed system.
This is because a parking site becomes accessible 
every time the previous busy time \HL{ends}.
In the next section, we investigate the relationship between $N_{b}$ and 
the distribution parameter $k$ of the exponential function.

\section{Analysis}
\subsection{First approximation level}
At the beginning of this study, we proposed a fundamental model, which 
is simulated by the concept of D-Fork system. 
By considering the effect of walking distance from the leftmost site, the 
occupied time $T_i$ of the $i-$th parking site can be modeled as follows:
\begin{eqnarray}
T_{i} &=& \tau_{s} + i \Delta l + \alpha \label{eq:mmc:ti}
\end{eqnarray}
Here, $\alpha$ is a constant parameter and $\Delta l$ is the distance between two parking sites.

\HL{In the right-hand side of Eq.(\ref{eq:mmc:ti}), the first term indicates the staying time in a parking site. 
	The second term indicates the traveling time to the parking site. In this approximation level, we ignore 
		the effect of volume exclusion effect in the second term; a particle is assumed to hop to 
		the neighboring cell per a step. Thus, the velocity of the particle becomes 1 
		since the length of a cell is set to 1. 
		That is why the notation of velocity does not emerge in the second term. 
		Instead, we introduce the $\alpha$ in the third term, assuming that the volume exclusion effect 
		can be approximated as constant values in the target system.}

\begin{figure}[t]
\begin{center}
\vspace{-7.0cm}
\includegraphics[width=\figscale\textwidth,clip, bb= 0 0 1500 1125]{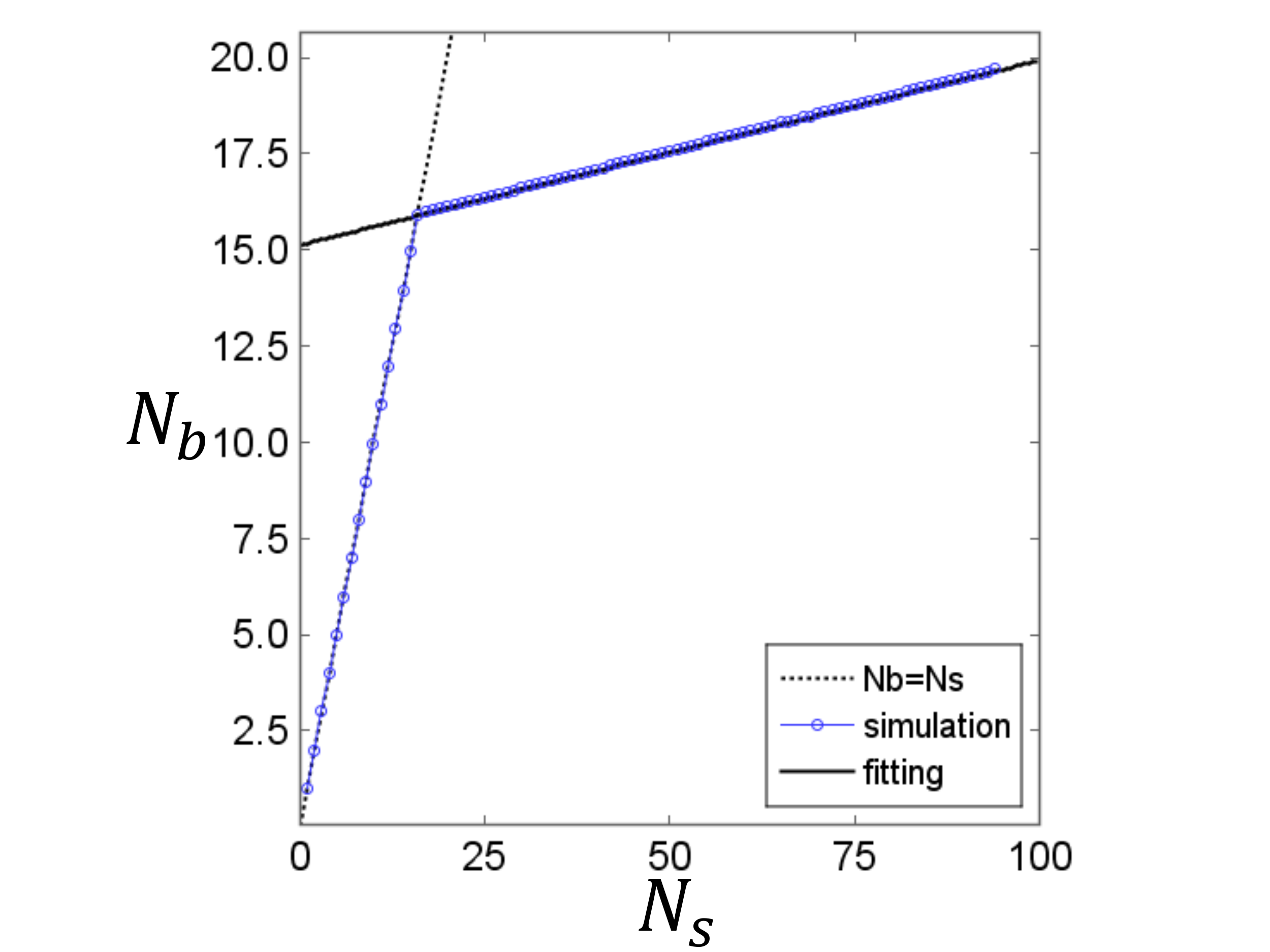}
\end{center}
\caption{Dependence of the number of busy sites $N_b$ on the total number of sites $N_{s}$ in the case of $k=0$.}
\label{fig:dependance-Ns}
\end{figure}

The maximum service rate $\mu_{max}$ and the minimum service rate $\mu_{min}$ 
\HL{are obtained by substituting $N_{s}$ and $1$ to Eq.(\ref{eq:mmc:ti}), as follows:}
\begin{eqnarray}
\mu_{max} &=& {T_{N_{s}}}^{-1} \label{eq:mmc:max} \\
\mu_{min} &=& {T_{1}}^{-1} \label{eq:mmc:min}
\end{eqnarray}
The averaged value of the service rate of the system $\mu_{avr}$ 
\HL{is obtained by calculating the arithmetic mean of Eq.(\ref{eq:mmc:ti}), as follows:}
\begin{eqnarray}
\mu_{avr} &=& \Biggl\{ \tau_{s} + \frac{1}{N_{s}}\sum_{i=1}^{N_s} (i \Delta l + \alpha ) \Biggr\}^{-1}\label{eq:mmc:avr1}\\
          &=& \Biggl\{ \tau_{s} + \frac{\Delta l}{2} N_{s} + \frac{\Delta l}{2} + \alpha  \Biggr\}^{-1} \label{eq:mmc:avr2}
\end{eqnarray}
\HL{Finally, the averaged number of busy sites $N_{b}$} is obtained 
by dividing Eq.(\ref{eq:mm1:lam}) by Eq.(\ref{eq:mmc:avr2}), as follows:
\begin{eqnarray}
\HL{N_{b}} &=& \frac{\lambda}{\mu_{avr}} \\ 
  &=& \Biggl(\frac{\Delta l}{2\tau_{in}}\Biggr) N_{s} + \frac{1}{\tau_{in}}\Biggl(\tau_{s} + \frac{\Delta l}{2} + \alpha \Biggr) \label{eq:mmc:Uavr}
\end{eqnarray}
Equation~(\ref{eq:mmc:Uavr}) shows that the number of busy sites $N_{b}$ is a linear function 
of the number of sites $N_{s}$. 
Figure~\ref{fig:dependance-Ns} shows the simulation result of the dependence 
of the number of busy sites $N_b$ on the total number of sites $N_{s}$, 
in the case of $k=0$. It was observed that the break in line occurs 
at around $N_{s} = 17$, which is because all the sites 
are in use due to the lack of capacity of sites \HL{when $N_{s} < 17$}. 
In comparison with Fig.\ref{fig:breakdown}, the y-intercept 
value of the fitting line in Fig.\ref{fig:dependance-Ns} 
corresponds to the value of the number of occupied sites $N_{o}$. 
This is because the increment of the number of busy sites $N_{b}$ 
depends only on the increase in the number of reserved sites $N_{r}$. 
By fitting the line at $N_{s} > 17$ according to Eq.(\ref{eq:mmc:Uavr}),
the parameter $\alpha$ is obtained to be 6.215 as a fitting result.
It was confirmed from \HL{the red colored line in Fig.\ref{fig:second-order}} that 
the simulation results are bounded between the maximum case (the dotted line) and the minimum case (the dashed line). 
\HL{Besides, } the averaged case of classical $M/M/c$ with the parameter $\alpha$ is found to semi-experimentally correspond to the case of $k = 0$.

\subsection{Second approximation level}
On the basis of an assumption that the site usage distributions 
obey the exponential function $k e^{-kx}$, 
we correct Eq.(\HL{\ref{eq:mmc:avr2}}) by replacing the arithmetic mean 
by the weighted average using the exponential function $k e^{-kx}$.
\HL{The number of busy sites $N_{b}$} is calculated as follows: 
\begin{eqnarray}
\HL{N_{b}} &=& \frac{\tau_{s}}{\tau_{in}} + \frac{1}{\tau_{in}} \frac{\sum_{i=1}^{N_{s}} (i \Delta l + \alpha) E_{i}}{\sum_{i=1}^{N_{s}} E_{i} } \label{eq:mmc:expo}\\ 
E_{i} &=& k\cdot{\rm exp}(-k \frac{i}{N_{s}})
\end{eqnarray}
\begin{figure}[t]
\begin{center}
\vspace{-7.0cm}
\includegraphics[width=\figscale\textwidth,clip, bb= 0 0 1500 1125]{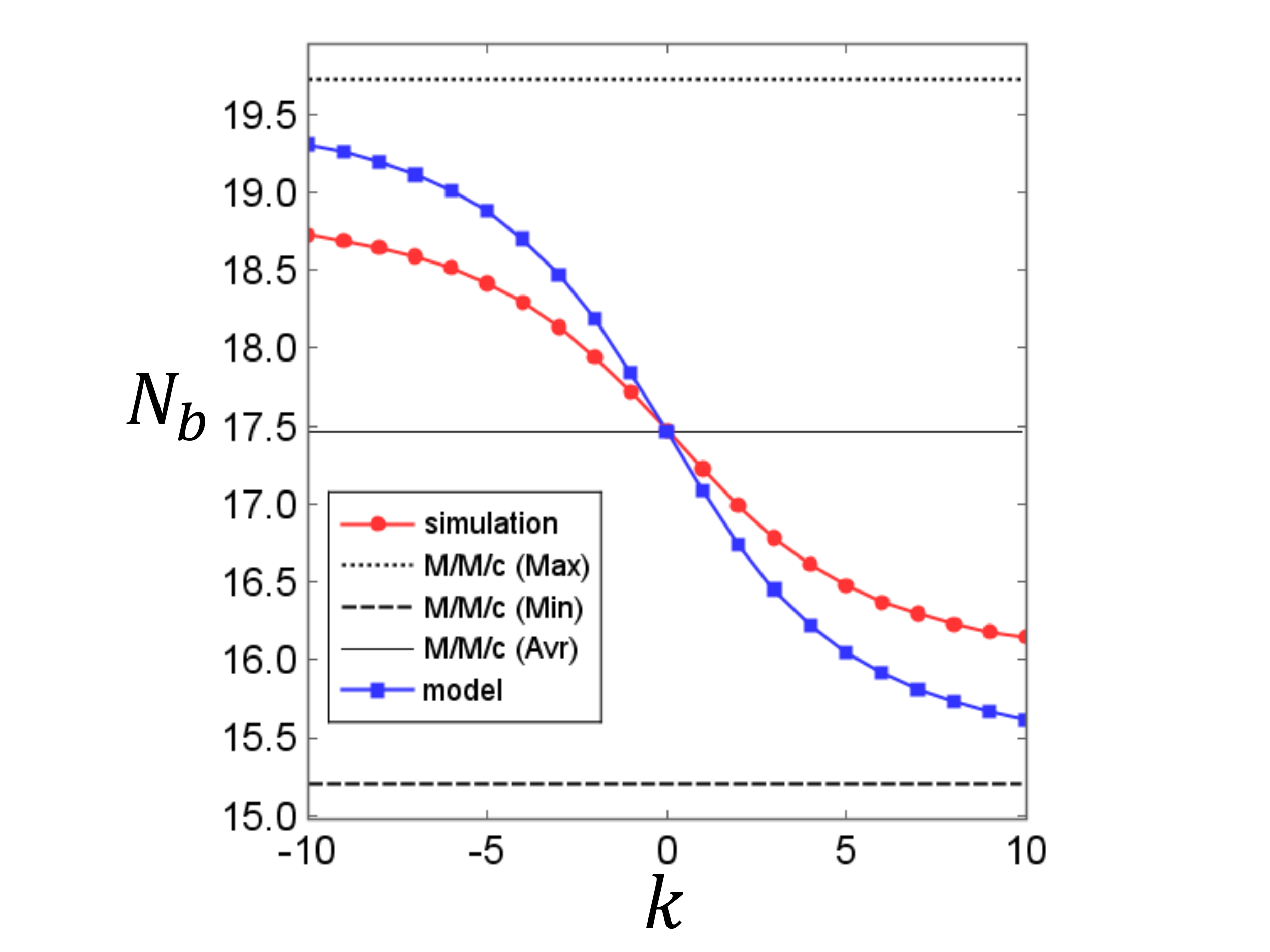}
\end{center}
\caption{\HL{Comparison of the simulation results, 
	the estimated values obtained using Eq.(\ref{eq:mmc:expo}), 
	the maximum values by dividing Eq.(\ref{eq:mm1:lam}) by $\mu_{max}$,
	the minimum values by dividing Eq.(\ref{eq:mm1:lam}) by $\mu_{min}$,
	the averaged values obtained by using Eq.(\ref{eq:mmc:avr2}), 
	for different values of distribution parameter $k$ between -10 and 10.}}
\label{fig:second-order}
\end{figure}

\HL{The red colored line and blue colored line in Fig.\ref{fig:second-order} show the comparison of simulation results 
	and the estimated values obtained using Eq.(\ref{eq:mmc:expo}), respectively.} 
It was confirmed that the feature of S-shape curve is observed in both simulations and approximations. 
However, the number of busy sites $N_{b}$ calculated by Eq.(\ref{eq:mmc:expo}) becomes overestimated/underestimated at the both sides of $k < 0$ and $k > 0$.

\begin{figure}[t]
\begin{center}
\vspace{-7.0cm}
\includegraphics[width=\figscale\textwidth,clip, bb= 0 0 1500 1116]{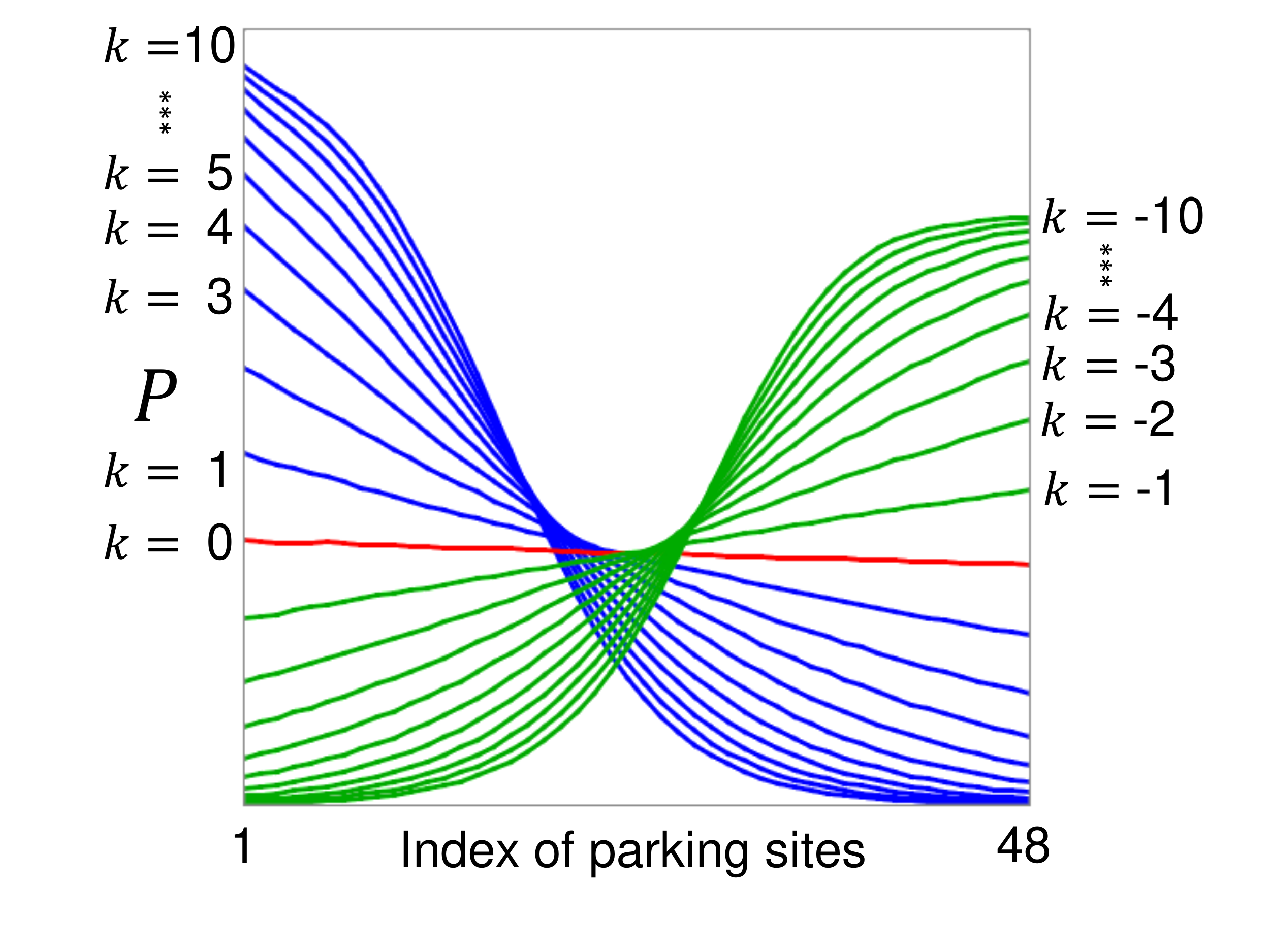}
\end{center}
\caption{All the distributions for different values of parameter $k$ between $-10$ and $10$.} 
\label{fig:site-dists}
\end{figure}

In order to clarify the reason for the deviation, the site usage distributions were investigated. 
Figure~\ref{fig:site-dists} shows all the distributions
for different values of parameter $k$ between $-10$ and $10$. 
Obviously, the shape of each distribution is different 
from that of the exponential function. 
It should be noted that the reason why the distribution gets slightly biased to the leftmost site in the case of $k=0$ 
is that the parking site, which is closer to the leftmost site has a higher turnover rate because of the shorter walking distance.

\subsection{Third approximation level}
\subsubsection{Birth-death process for walking direction}\label{seq:birthdeathp}
In this section, we describe the site usage distributions 
	by introducing the birth-death process for walking direction.
Namely, the particles in transport before stopping at a site on the 
parking lane (red colored particles in Fig.\ref{fig:schemview}) 
are regarded as ``surviving''. On the contrary, an event in which
a particle stops at a site indicates the ``death'' of the particle. 

We define a random variable $X$, which donates a position on the traveling lane, 
   $f(x)$ is defined as the probability density function of $X$. 
From these definitions, the cumulative density function $F(x)$ of $f(x)$ 
can be expressed as follows:
\begin{eqnarray}
F(x) &:=& P(X \le x) = \int_{0}^{x} f(x) dx \label{eq:sd:cdf}
\end{eqnarray}
$F(x)$ represents the probability of ``death'' at position $x$.
Conversely, we define $S(x)$ as the probability of surviving at position $x$, as follows:
\begin{eqnarray}
S(x) &:=& 1- F(x) = P(X > x) \label{eq:sd:scdf}
\end{eqnarray}
Besides, a hazard function $h(x)$ is defined as follows:
\begin{eqnarray}
h(x) &:=& \lim_{\Delta \to 0} \frac{1}{\Delta}\cdot P(x < X < x + \Delta|X > x) \label{eq:sd:hasf} 
\end{eqnarray}
$h(x)$ is the probability density that a particle stops at a site at the position between $x$ and $x+\Delta$.
Equation.(\ref{eq:sd:hasf}) can be transformed as follows:
\begin{eqnarray}
      &=& \lim_{\Delta \to 0} \frac{1}{\Delta}\cdot \frac{P\{(x < X < x + \Delta) \cap (X > x)\} }{P(X > x)} \\
      &=& \lim_{\Delta \to 0} \frac{1}{\Delta}\cdot \frac{P(x < X < x + \Delta)}{P(X > x)} 
\end{eqnarray}
From Eq.(\ref{eq:sd:cdf}) and Eq.(\ref{eq:sd:scdf}), 
\begin{eqnarray}
~~~~~~&=& \frac{1}{P(X > x)}\lim_{\Delta \to 0}\frac{F(x+\Delta)-F(x)}{\Delta} \\
~~~~~~&=& \frac{F'(x)}{S(x)} \\
~~~~~~&=& -\frac{S'(x)}{S(x)} = -\frac{d}{dx}({\rm log}(S(x))) \label{eq:sd:log}
\end{eqnarray}
Here, we impose an initial condition $S(0)=1$ to Eq.(\ref{eq:sd:log}), since 
the probability that a particle survives at the position of zero always becomes $1$.
Then, the differential equation Eq.(\ref{eq:sd:log}) can be soloved as follows:
\begin{eqnarray}
S(x) &=& {\rm exp}(-H(x)) \label{eq:S:res}
\end{eqnarray}
Here, we introduce the $H(x)$, as follows:
\begin{eqnarray}
H(x) &:=& \int_{0}^{x}h(x)dx
\end{eqnarray}
We obtain $F(x)$ and $f(x)$ from Eq.(\ref{eq:S:res}) as follows:
\begin{eqnarray}
F(x) &=& 1 - {\rm exp}(-H(x)) \label{eq:F:res} \\
f(x) &=& h(x)\cdot{\rm exp}(-H(x))  \label{eq:sf:res}
\end{eqnarray}
$f(x)$ corresponds to the probability distribution of site usage 
since $F(x)$ is the probability of death at position $x$.
We have successfully obtained a general formula for 
site usage distribution in the proposed system.
We introduce extreme statistics in Section~\ref{sec:histexstat2} to determine the specific formula of $H(x)$.

\subsubsection{Introduction of order statistics\\}\label{sec:aproxestat1} 
\HL{The derivation of Eq.(\ref{eq:F:res}) lacks the information of order statistics of the random variable $X$. 
	We introduce the concept of order statistics to the proposed system in this section, 
	   as a preliminary work for the approximation by extreme statistics in Section~\ref{sec:histexstat2}. 
}

\HL{Let us consider the situation that a single particle is inserted from the queue to the leftmost site at certain intervals of arrival time
	during the total $n$ time steps. We name the $i-$th inserted particle to the leftmost site simply as ``i-th particle''.
	We define the random variables $X_{i}$, which indicates the position that $i-$th particle stops at during its travel.
	As similarly in the previous section, the $i-$th particle is judged as ``death'' when $X_{i} \le x$. 
	On the contrary, the particle is judged as ``surviving'' when $X_{i} > x$.}

\HL{An identical cumulative density function $A(x)$ of ${X_{1}, X_{2},\dots, X_{n}}$ is expressed, as follows:}
\begin{eqnarray}
A(x) &:=& P(X_{i}\le x),~~(i=1,2,\dots, n) \label{eq:Ax}
\end{eqnarray}

\begin{figure}[t]
\begin{center}
\vspace{-5.0cm}
\includegraphics[width=\figscale\textwidth,clip, bb= 0 0 2002 1127]{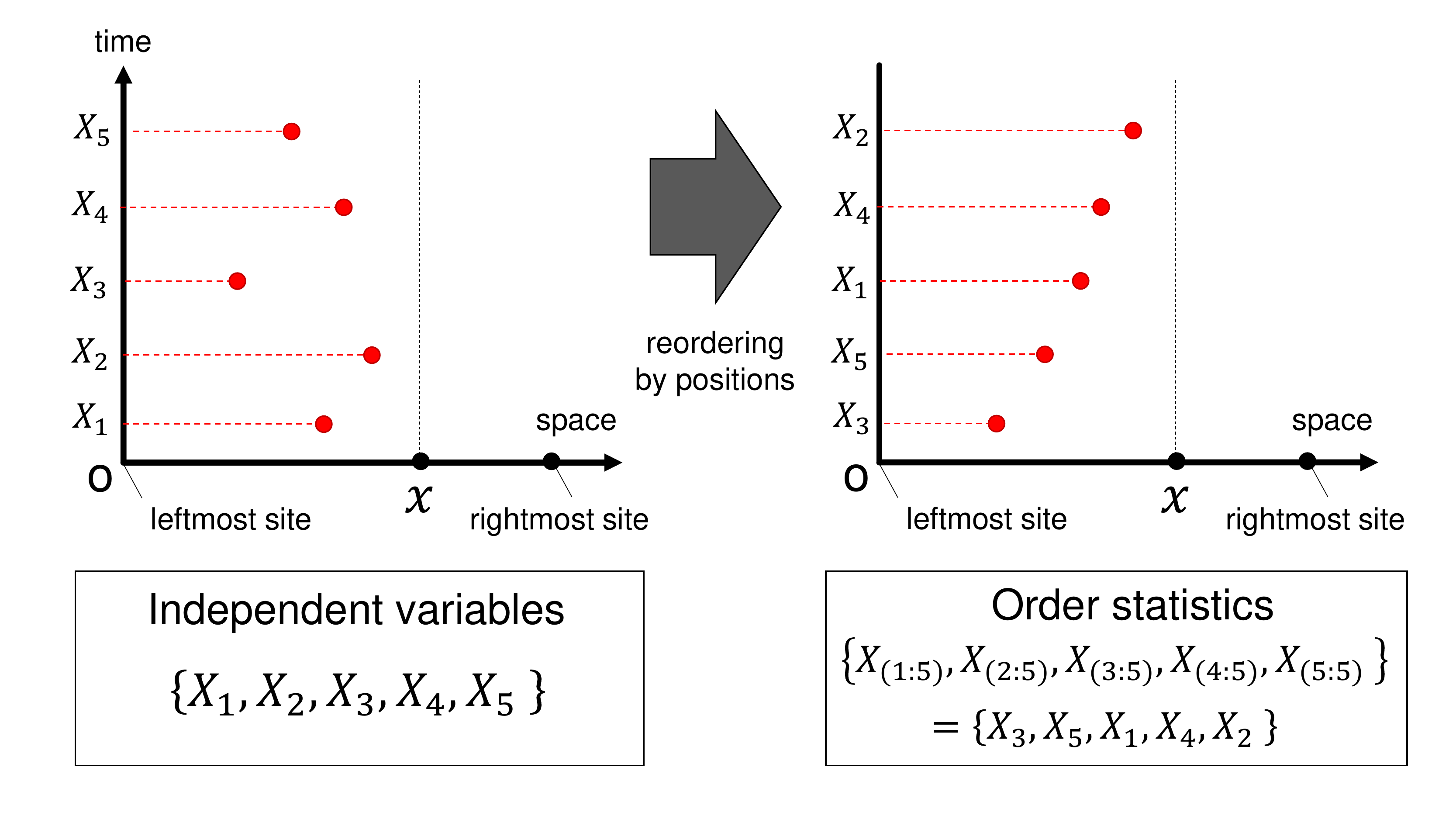}
\end{center}
\caption{\HL{Relationship between the independent variables and the order statistics in case of $n=5$.} }
\label{fig:relivandos:n5}
\end{figure}

\HL{The order statistics of ${X_{1}, X_{2},\dots, X_{n}}$, which is obtained by rearranging the ${X_{1}, X_{2},\dots, X_{n}}$
	in an ascending order, is represented as follows:}
\begin{eqnarray}
&X_{(1:n)} \le X_{(2:n)} \le \dots \le X_{(n:n)}&
\end{eqnarray}
\HL{From the definition of the order statistics, it is obvious that 
	the $i-$th order statistic $X_{(i:n)}$ corresponds to the $i-$th largest original independent variables.
	The relationship between the original independent variables and the order statistics in case of $n = 5$ is depicted in Fig.\ref{fig:relivandos:n5}.}

\HL{A cumulative density function $A_{X_{(m:n)}}(x) $ of the $m-$th statistic $X_{(m:n)}$ is defined as follows:
\begin{eqnarray}
	A_{X_{(m:n)}}(x) &:=& P(X_{(m:n)}\le x) \label{eq:mthcdf}
\end{eqnarray}
}
\HL{Considering the fact that $X_{(i:n)}$ corresponds to the $i-$th largest original independent variables, it can be said that 
	Eq.(\ref{eq:mthcdf}) indicates the probability that at least $m$ variables of
	   	${X_{1}, X_{2},\dots, X_{n}}$ become equal or less than the position $x$ (= the state of ``death'').}
	
\HL{By using the probability $P_{j}$ that exactly $j$ variables of ${X_{1}, X_{2},\dots, X_{n}}$ 
	becomes equal or less than the position $x$, the right-hand side of Eq.(\ref{eq:mthcdf}) is represented as follows:
\begin{eqnarray}
	P(X_{(m:n)}\le x) &=& \sum_{j=m}^{n} P_{j} \label{eq:mthcdf2}
\end{eqnarray}
}
\HL{The probability $P_{i}$ is further decomposed as follows;  
	there are ${}_n \mathrm{C} _j$ different combinations of $j$ variables from ${X_{1}, X_{2},\dots, X_{n}}$.
In each case, $j$ variables become ``death'' with the probability $A(x)$ 
	and that of $n-j$ variables become ``survival'' with the probability $1-A(x)$; 
	the cumulative density function $A_{X_{(m:n)}}(x)$ is represented by using Eq.(\ref{eq:mthcdf}) and Eq.(\ref{eq:mthcdf2}), as follows:
\begin{eqnarray}
A_{X_{(m:n)}}(x) &=& \sum_{j=m}^{n} {}_n \mathrm{C} _j~A(x)^{j}(1-A(x))^{n-j} \label{eq:orderstat3}
\end{eqnarray}
}
\HL{Now we obtain the precise expression of Eq.(\ref{eq:mthcdf}). 
Equation~(\ref{eq:orderstat3}) includes all the possible patterns of particle arrivals for the time direction  
during the total $n$ time steps.}

\HL{Unfortunately}, it is difficult to derive the probability distribution of site usage directly from Eq.(\ref{eq:orderstat3})
because the cumulative density function $A(x)$ at each time step is unknown.
To solve this problem, we propose to approximate the site usage distribution 
by the asymptotic distribution of the distribution of extreme order statistics in the next section.

\subsubsection{Approximation by extreme statistics}\label{sec:histexstat2}
The maximum order statistics $Z_{n}$ and 
the minimum order statistics $Y_{n}$ are defined, respectively, as follows:
\begin{eqnarray}
Z_{n} &:=& X_{(n:n)} = {\rm max}\{{X_{1}, X_{2},\dots, X_{n}}\} = \argmax_{1 \leq i \leq n} X_{i} \label{eq:extstat1} \\
Y_{n} &:=& X_{(1:\HL{n})} = {\rm min}\{{X_{1}, X_{2},\dots, X_{n}}\} = \argmin_{1 \leq i \leq n} X_{i} \label{eq:extstat2}
\end{eqnarray}

In this paper, we propose to approximate the cumulative probability distribution of site usage in Eq.(\ref{eq:sd:cdf}) 
by the distribution of extreme order statistics. 
Here, we have two candidates of $P(Z_{n} \le x)$ and $P(Y_{n} \le x)$. 
	
\begin{figure}[t]
\begin{center}
\vspace{-3.5cm}
\includegraphics[width=\figscale\textwidth,clip, bb= 0 0 1972 836]{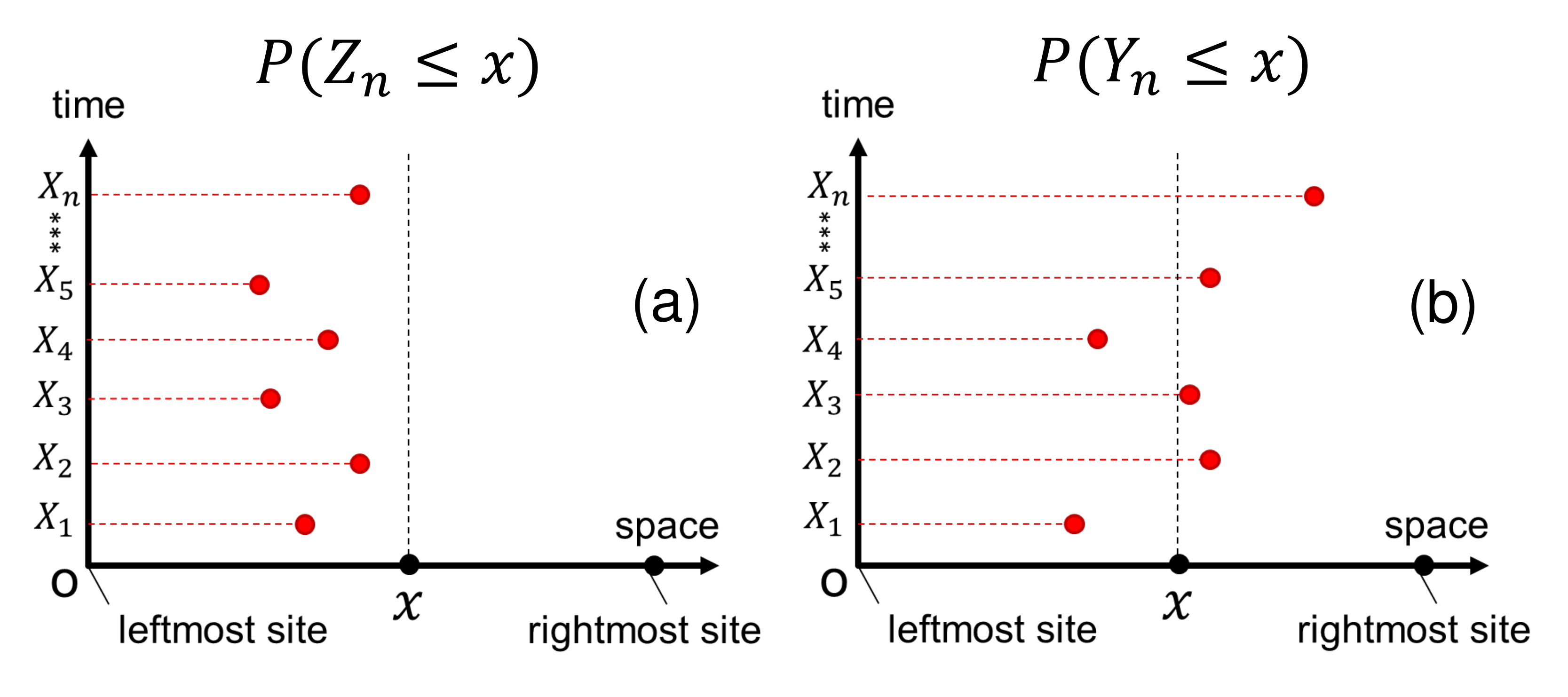}
\end{center}
\caption{Schematic view of the distributions of two different extreme order statistics.} 
\label{fig:maxminskemview}
\end{figure}

\HL{We are able to say} that the selection of $P(Y_{n} \le x)$ is appropriate for the approximation of Eq.(\ref{eq:sd:cdf}) 
considering the physical meaning of these extreme order statistics. 
$P(Z_{n} \le x)$ describes the probability that not a single $X_{i}$ becomes
larger than the position $x$ during the time step $n$, as shown in Fig.\ref{fig:maxminskemview}(a).
Because the situation depicted in Fig.\ref{fig:maxminskemview}(a) seldom 
occurs in the proposed system, replacing the random variables $X$ in Eq.(\ref{eq:sd:cdf}) by the 
maximum order statistics of $Z_{n}$ is not appropriate.
On the contrary, as shown in Fig.\ref{fig:maxminskemview}(b), $P(Y_{n} \le x)$ 
describes the probability that at least one of $X_{i}$ becomes 
smaller than the position $x$ during time step $n$; 
therefore, the asymptotic distribution of minimum order statistics of $Y_{n}$ 
is suitable for describing the behaviors of 
the proposed system, compared to the former case. 

\HL{We approximate the cumulative density distribution of site usage in Eq.(\ref{eq:sd:cdf}) 
by the distribution of minimum order statistics $Y_{n}$, as follows: 
\begin{eqnarray}
F(x) &\approx&  P(Y_{n} \le x) \label{eq:fandp} 
\end{eqnarray}
}
\HL{For adequate large $n$, it is known that the distribution of 
	minimum order statistics $Y_{n}$ asymptotic to the following extreme value distribution $M(x)$ 
in case that the random variables of ${X_{1}, X_{2},\dots, X_{n}}$ follow exponential distributions\cite{d18a0a18-8367-4589-8f58-7f91500d41a8, doi:10.1201/9781420087444.ch1}:
\begin{eqnarray}
 P(Y_{n} \le x) &\rightarrow& M\Biggl(\frac{x-\tilde{c}}{\tilde{b}}\Biggr) \\
	            &=& 1 - {\rm exp}\Biggl[-{\rm exp}\Biggl(\frac{kx-c}{b}\Biggl)\Biggr] \label{eq:fandw}
\end{eqnarray}
Here, ($\tilde{c}$, $\tilde{b}$) are normalizing constants, which are selected to convert the location and scale 
so that the extreme value distribution $M$ does not diverge and degenerate. 
The $(c, b)$ is $(k\tilde{c}, k\tilde{b})$, respectively.
For detail description on the derivation of Eq.(\ref{eq:fandw}), see the Appendix.

Now we obtain the distribution $F(x)$ of the proposed system, as follows:
\begin{eqnarray}
F(x) &=& 1 - {\rm exp}\Biggl[-{\rm exp}\Biggl(\frac{kx-c}{b}\Biggl)\Biggr]   \label{eq:F:fin}
\end{eqnarray}
}
The probability density function $f(x)$ is obtained as follows:
\HL{
\begin{eqnarray}
f(x) &=& \frac{k}{b}\cdot{\rm exp}\Biggl(\frac{kx-c}{b}\Biggl)\cdot{\rm exp}\Biggl[-{\rm exp}\Biggl(\frac{kx-c}{b}\Biggl)\Biggr] \label{eq:sf:fin}
\end{eqnarray}
}
From Eq.(\ref{eq:sf:fin}), the cumulative hazard function $H(x)$ is found to 
become an exponential function:
\HL{
\begin{eqnarray}
H(x) &=& {\rm exp}\Biggl(\frac{kx-c}{b}\Biggl)
\end{eqnarray}
}

The selection of $Y_{n}$ is validated from the point of mathematical derivation.
If we select $Z_{n}$, the right-hand side of Eq.(\ref{eq:F:fin}) becomes ${\rm exp}[-{\rm e}^{-(kx-f)/g}]$.
This description contradicts with the formula obtained in Eq.(\ref{eq:F:res}). 
In this case, the relationship between Eq.(\ref{eq:F:res}) and Eq.(\ref{eq:sf:res}) is not satisfied.

\HL{It is not easy to mathematically derive the constant parameters of $(c,b)$ 
of minimum order statistics $Y_{n}$,
we determine these parameters by fitting the simulation results in the next section.}

\subsubsection{Corrections of the $M/M/c$ queueing model\\}\label{sec:aproxestat} 
\begin{figure}[t]
\begin{center}
\vspace{-7.0cm}
\includegraphics[width=\figscale\textwidth,clip, bb= 0 0 1500 1108]{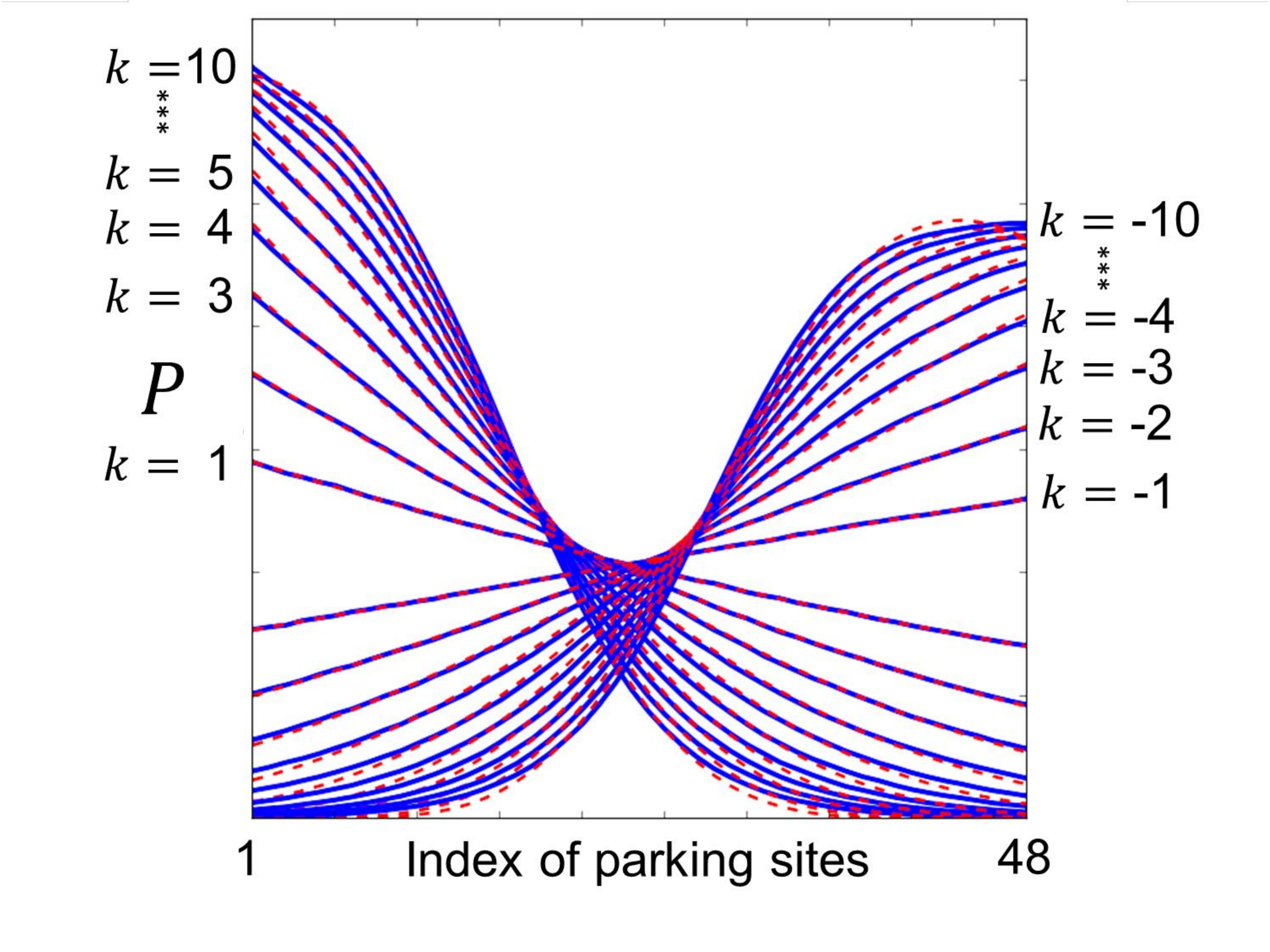}
\end{center}
\caption{All the distributions fitted by Eq.(\ref{eq:sf:fit}) 
	for different values of parameter $k$ between $-10$ and $10$.} 
\label{fig:fitting}
\end{figure}

\begin{figure}[t]
\begin{center}
\vspace{-7.0cm}
\includegraphics[width=\figscale\textwidth,clip, bb= 0 0 1530 1120]{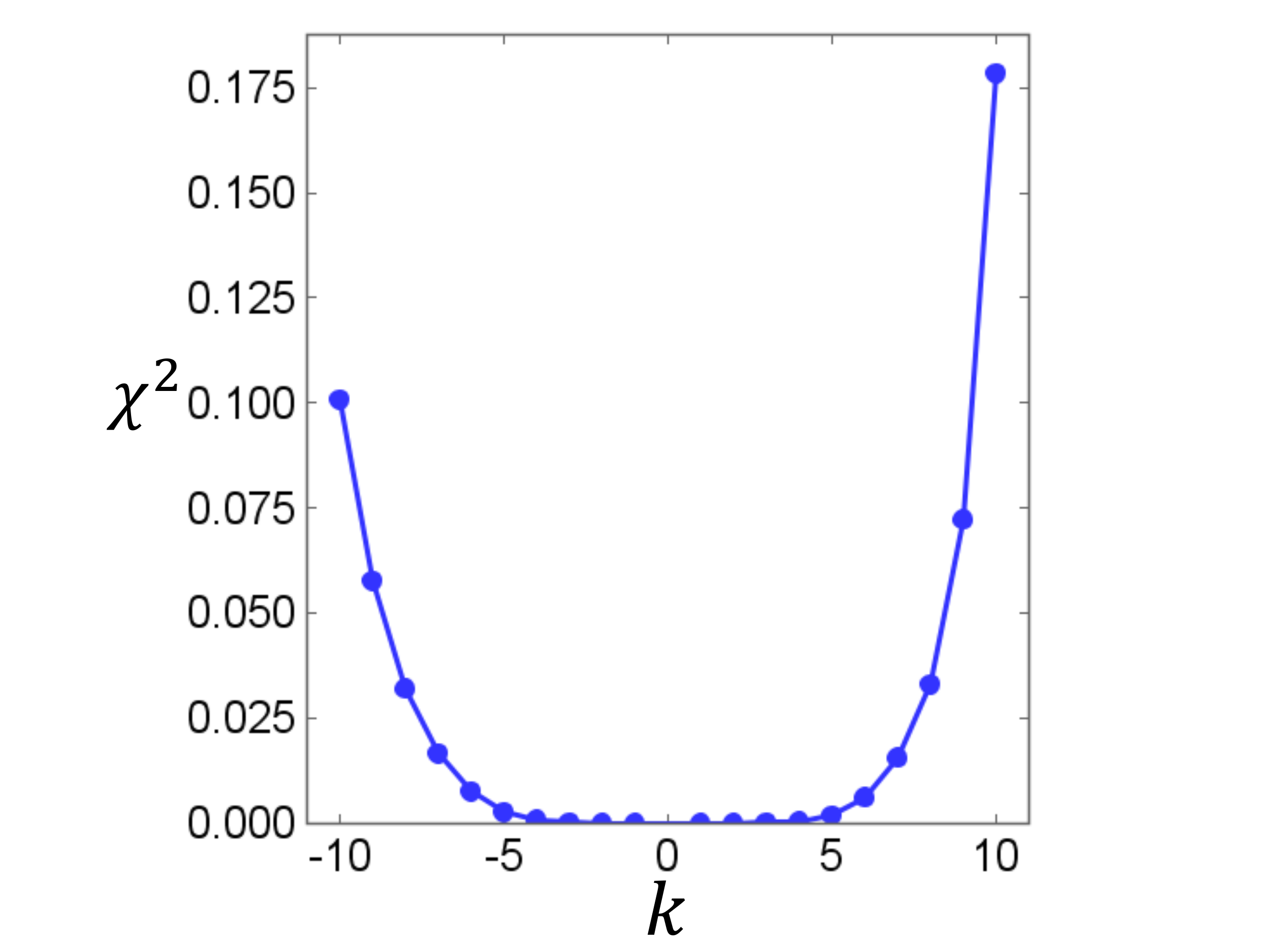}
\end{center}
\caption{Dependence of the chi-square of fitting results in Fig.\ref{fig:fitting} on the different values of parameter $k$.} 
\label{fig:kaisq}
\end{figure}

Let us get back to the subject of queueing theory. 
We attempt to correct the weighted calculation in Eq.(\ref{eq:mmc:expo}) by replacing the exponential function by 
the fitting function of the simulation results. 
We adopt Eq.(\ref{eq:sf:fin}) as the fitting function, \HL{admitting 
the transformation of the scale of Eq.(\ref{eq:sf:fin}) by 
using constant parameter $a$, as follows:}
\HL{
\begin{eqnarray}
f_{\rm {FIT}}(x) &=& a\frac{k}{b}\cdot{\rm exp}\Biggl(\frac{kx-c}{b}\Biggl)\cdot{\rm exp}\Biggl[-{\rm exp}\Biggl(\frac{kx-c}{b}\Biggl)\Biggr] \label{eq:sf:fit}
\end{eqnarray}
}
Figure~\ref{fig:fitting} shows all the cases of exponential 
distributions fitted by Eq.(\ref{eq:sf:fit}) 
for different values of parameter $k$ between $-10$ and $10$. 
The dashed red colored lines indicate fitting results by the least squares method.
Figure~\ref{fig:kaisq} shows the dependence of the chi-square 
of fitting results in Fig.\ref{fig:fitting} on the different values of parameter $k$.
Obviously, in Fig.\ref{fig:kaisq}, it was observerd that the accuracy of curve fitting deteriorates as the bias to the right/left side increases. 
The reason for this is interpreted as follows. 
As the bias to the right/left side increases, congestion occurs 
in the neighboring area of the rightmost/leftmost site.
Because the effect of congestion is not considered in the deviation of Eq.(\ref{eq:sf:fit}), the difference at both sides of the edges emerges. 

We correct the weighted calculations of the proposed queueing model in Eq.(\ref{eq:mmc:expo}) by replacing
the weighted function with the function in Eq.(\ref{eq:sf:fit}), as follows:
\begin{eqnarray} 
\HL{N_{b}} &=& \frac{\tau_{s}}{\tau_{in}} + \frac{1}{\tau_{in}} \frac{\sum_{i=1}^{N_{s}} (i \Delta l + \alpha) E_{i}}{\sum_{i=1}^{N_{s}} E_{i} } \label{eq:mmc:modif}\\ 
E_{i} &:=& f_{\rm {\HL{FIT}}}(\frac{i}{N_{s}})
\end{eqnarray} 

\begin{figure}[t]
\begin{center}
\vspace{-7.0cm}
\includegraphics[width=\figscale\textwidth,clip, bb= 0 0 1500 1125]{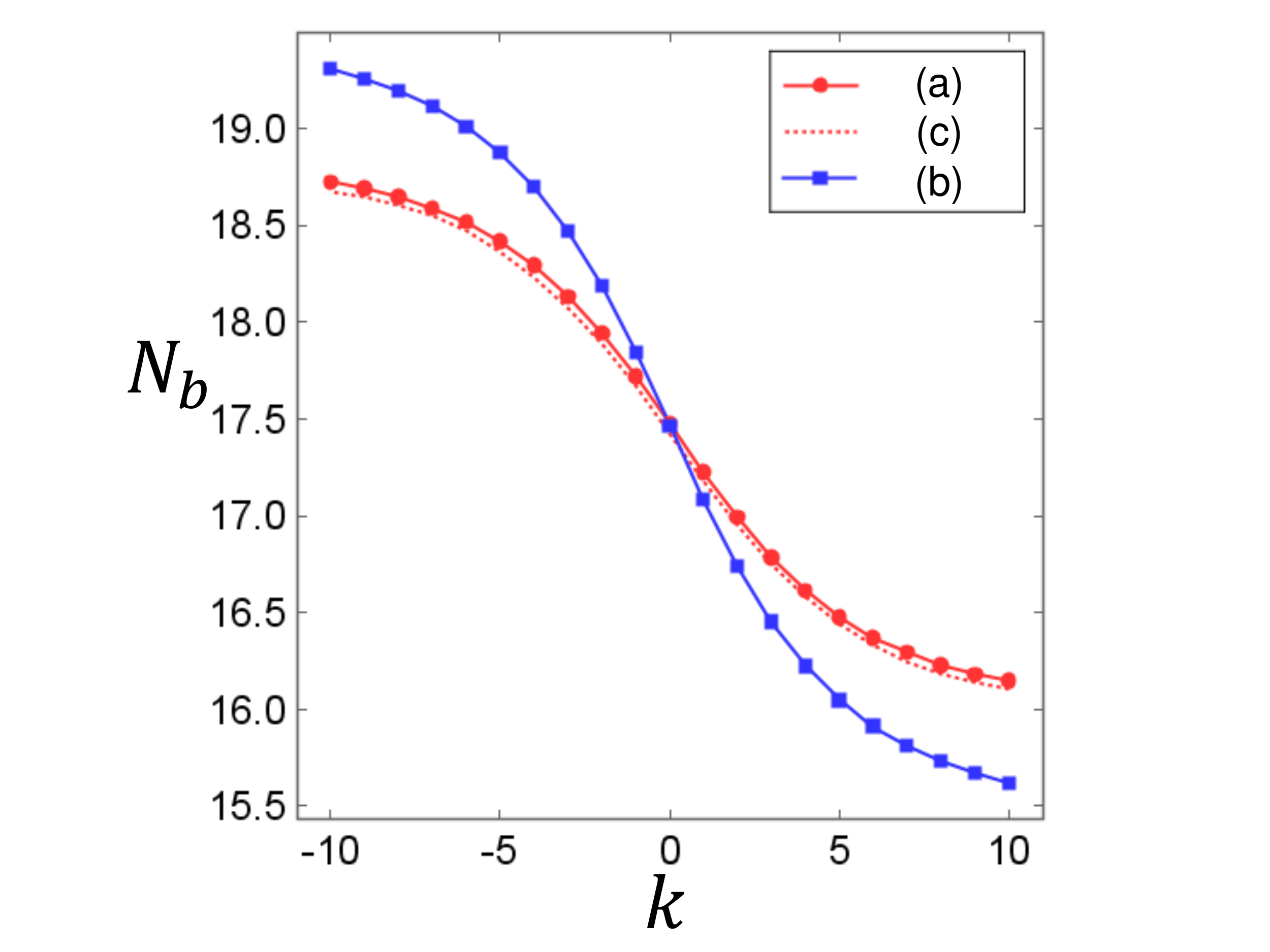}
\end{center}
\caption{
(a) the simulation results in Fig.\ref{fig:second-order},
(b) second-order proposed model exhibited in Eq.(\ref{eq:mmc:expo}), and
   (c) third-order proposed model exhibited in Eq.(\ref{eq:mmc:modif}). }
\label{fig:compare:threemmc}
\end{figure}

Figure~\ref{fig:compare:threemmc} shows a comparison of (a) the simulation results 
in Fig.\ref{fig:second-order} with (b) the estimated values obtaind using 
Eq.(\ref{eq:mmc:expo}) and (c) the estimated values obtained using Eq.(\ref{eq:mmc:modif}).
It was confirmed that our proposed model shows a good agreement with the 
simulation results compared to the model exhibited in Eq.(\ref{eq:mmc:expo}).
This result indicates that the proposed method, which estimates the service rate $\mu_{s}$ by using weighted calculations of the site usage distributions, 
is an effective approach under certain conditions.

\section{Conclusion}
We introduced a totally asymmetric simple exclusion process on a traveling lane
equipped with a queueing system and functions of site assignments along the parking lane. 
In this study, we investigated the relationship between the utilization 
of parking sites and the effect of site assignments in the proposed system. 
The contributions of this study are as follows:

We proposed an approximation model to describe the 
site usage distributions of the proposed system 
on the basis of birth-death process for the spatial direction 
and extreme statistics for the time direction.
The specific formula in the case where the random variables 
follow exponential distributions are described.
In addition, our proposed $M/M/c$ queueing model, whose service rate 
is determined by the weighted calculation of site usage distributions, 
   shows good agreement with the simualtion results. 

As mentioned in the introduction, the major scope of the current research 
is to describe the relationship between the utilization of parking sites 
and the effect of site assignments in the proposed system. 
Accordingly, we obtained insightful results from the findings of this study.

\section*{Acknowledgements}
This research was supported by MEXT as ``Post-K Computer Exploratory Challenges'' (Exploratory Challenge 2: Construction of Models for Interaction Among Multiple Socioeconomic Phenomena, Model Development and its Applications for Enabling Robust and Optimized Social Transportation Systems)(Project ID: hp180188), partly supported by JSPS KAKENHI Grant Numbers 25287026 and 15K17583.
We would like to thank Editage (www.editage.jp) for English language editing.

\appendix

\HL{
\section{Asymptotic distributions of the distributions of extreme order statistics}\label{apend:asym}
\HL{The distributions of maximum order statistics $Z_{n}$ and 
	minimum order statistics $Y_{n}$ are represented as follows:
\begin{eqnarray}
P(Z_{n} \le x) &=& P(X_{(n:n)} \le x) = A_{X_{(n:n)}}(x) \label{dmaxostat}\\
P(Y_{n} \le x) &=& P(X_{(1:n)} \le x) = A_{X_{(1:n)}}(x) \label{dminostat}
\end{eqnarray}
}
\HL{For adequate large $n$, the distributions of these two extreme order statistics 
are assumed to asymptotic to the extreme value distributions, respectively, as follows:
\begin{eqnarray}
P(Z_{n} \le x)  &\rightarrow& G\Biggl(\frac{x-{a}_{n}}{{b}_{n}}\Biggr) \label{eq:sympzn}\\ 
P(Y_{n} \le x)  &\rightarrow& M\Biggl(\frac{x-{c}_{n}}{{d}_{n}}\Biggr) \label{eq:sympyn}
\end{eqnarray}
}
\HL{Here, (${a}_{n}$, ${b}_{n}$) are normalizing constants, which are selected to convert 
the location and scale of $G$ so that the extreme value distribution $G$ does not diverge and degenerate.
The same is true of (${c}_{n}$, ${d}_{n}$).
The assumption of the existance of $G$ and Eq.(\ref{eq:sympzn}) are validated on condition that Eq.(\ref{eq:extstat:cond}) is satisfied
\cite{d18a0a18-8367-4589-8f58-7f91500d41a8, doi:10.1201/9781420087444.ch1}.
If they are validated, the asymptotic distribution $M(x)$ of minimum order statistics $Y_{n}$ is obtained 
from the following relationship:
\begin{eqnarray}
M(x) &=& 1-G(-x) \label{eq:relgandw}
\end{eqnarray}
}
}
\section{Trinity Theorem}\label{apend:tt}
A population distribution $F$ is assumed to belong to 
a domain of attraction of an extreme value distribution $G$; this assumption is denoted as $F \in \mathcal{D}(G)$.
R.A. Fisher and L.H.C Tippett\cite{fisher_tippett_1928} mathematically proved the following relationship for maximum order statistics $Z_{n}$:
\begin{eqnarray}
 F \in \mathcal{D}(G) \nonumber \\ 
	  \Leftrightarrow \lim_{n \to \infty}F^{\HL{n}}(a_{n}x+b_{n}) &=& G(x),~~a_{n} > 0, b_{n} \in \mathbb{R}\label{eq:extstat:cond} 
\end{eqnarray}
After considerable efforts, mathematicians
${\rm Fr{\rm \acute{e}}chet}$\cite{Frechet_Maurice_Sur_1928}, R. A. Fisher and L.H.C Tippett\cite{fisher_tippett_1928}, and Gnedenko\cite{Gnedenko} 
proved a notable fact that only three types of extreme distributions 
exist, which are as follows:
\begin{eqnarray}
{\rm Gumbel} : G(x) &=& {\rm exp}[-{\rm exp}(-x)],~~x\in \mathbb{R} 
\label{eq:extstat:ttt1}\\
{\rm Fr{\rm \acute{e}}chet} : G(x) &=& {\rm exp}(-x^{-\alpha}),~~x\ge 0,~\alpha > 0 
\label{eq:extstat:ttt2}\\
{\rm Weibull} : G(x) &=& {\rm exp}[-(-x)^{\alpha}],~~x\le 0, \alpha \ge 0 
\label{eq:extstat:ttt3}
\end{eqnarray}
The series of equations from Eq.(\ref{eq:extstat:ttt1}) to 
Eq.(\ref{eq:extstat:ttt3}) is called {Trinity Theorem}, which 
indicates that any population distribution $F$ is asymptotic to 
one of the three kinds of extreme distributions listed 
from Eq.(\ref{eq:extstat:ttt1}) to Eq.(\ref{eq:extstat:ttt3}), on the 
condition that the relation $F \in \mathcal{D}(G)$ is satisfied.

\section{The extreme value distributions of an exponential distribution}\label{apend:ee}
The asymptotic distribution for the case in which the random variables of 
${X_{1}, X_{2},\dots, X_{n}}$ follow exponential distributions is obtained, as follows.
A cumulative exponential function is written as follows:
\begin{eqnarray}
F(x) &:=& 1-{\rm exp}({-kx})
\end{eqnarray}
Here, we use the following indentity equation:
\begin{eqnarray}
F^{n}(a_{n}x+b_{n}) &=& \Biggl\{ 1+\frac{-n[1-F(a_{n}x+b_{n})]}{n}\Biggr\}^{n} \label{eq:extstat:exp1}
\end{eqnarray}
By selecting $a_{n} =1$ and $b_{n}=k^{-1}{\rm log} (n)$,
\begin{eqnarray}
-n[1-F(a_{n}x+b_{n})] &=& -{\rm exp}({-kx}),~~x \ge 0. \label{eq:extstat:exp2}
\end{eqnarray}
By substituting Eq.(\ref{eq:extstat:exp2}) into Eq.(\ref{eq:extstat:exp1}),
\begin{eqnarray}
\lim_{n \to \infty}F^{\HL{n}}(a_{n}x+b_{n}) &=& \lim_{n \to \infty} \Biggl( 1 + \frac{-{\rm e}^{-kx}}{n}\Biggr)^{n} \\
		&=& {\rm exp}[-{\rm exp}(-kx)] \label{eq:extstat:exp4}
\end{eqnarray}
\HL{From Eq.(\ref{eq:extstat:cond}), we obtain the expression of $G(x)$, as follows:
\begin{eqnarray}
G(x) &=& {\rm exp}[-{\rm exp}(-kx)]
\end{eqnarray}
Equation~(\ref{eq:extstat:exp4}) indicates that the asymptotic distribution $G(x)$ of maximum order statistics $Z_{n}$, when 
the random variables ${X_{1}, X_{2},\dots, X_{n}}$ follow an exponential distribution, 
	belongs to the family of Eq.(\ref{eq:extstat:ttt1}) in the Trinity Theorem.
}

\HL{From the relationship in Eq.(\ref{eq:relgandw}), we obtain the expression of $M(x)$, as follows:
\begin{eqnarray}
M(x) &=& 1 - {\rm exp}[-{\rm exp}(kx)] \label{eq:prc:w}
\end{eqnarray}
}
\HL{By substituting Eq.(\ref{eq:prc:w}) into Eq.(\ref{eq:sympyn}), we obtain as follows:
\begin{eqnarray}
P(Y_{n} \le x) &\rightarrow& M\Biggl(\frac{x-{c}_{n}}{{d}_{n}}\Biggr) \\
	            &=& 1 - {\rm exp}\Biggl[-{\rm exp}\Biggl(\frac{kx-\tilde{c}_{n}}{\tilde{d}_{n}}\Biggl)\Biggr] 
\end{eqnarray}
Here, $\tilde{c}_{n}$ and $\tilde{d}_{n}$ are $k{c}_{n}$ and $k{d}_{n}$, respectively.
}
\bibliographystyle{h-physrev3}
\bibliography{reference}

\clearpage




\end{document}